\documentclass[prl,aps,twocolumn,superscriptaddress]{revtex4}
\usepackage{amssymb}
\usepackage{epsfig}

\begin{document}

\title{Spin Texture in a Cold Exciton Gas}
\author{A.A.~High}
\affiliation{Department of Physics, University of California at San Diego, La Jolla, CA 92093-0319, USA}
\author{A.T.~Hammack}
\affiliation{Department of Physics, University of California at San Diego, La Jolla, CA 92093-0319, USA}
\author{J.R.~Leonard}
\affiliation{Department of Physics, University of California at San Diego, La Jolla, CA 92093-0319, USA}
\author{Sen Yang}
\affiliation{Department of Physics, University of California at San Diego, La Jolla, CA 92093-0319, USA}
\author{L.V.~Butov}
\affiliation{Department of Physics, University of California at San Diego, La Jolla, CA 92093-0319, USA}
\author{T. Ostatnick\'{y}}
\affiliation{Faculty of Mathematics and Physics, Charles University in Prague, Ke Karlovu 3, 121 16 Prague, Czech Republic}
\author{A.V.~Kavokin}
\affiliation{School of Physics and Astronomy, University of Southampton, SO17 1BJ, Southampton, United Kingdom}
\author{A.C.~Gossard}
\affiliation{Materials Department, University of California at Santa Barbara, Santa Barbara, CA 93106-5050, USA}
\date{\today}

\begin{abstract}
\noindent We report on the observation of a spin texture in a cold exciton gas in a GaAs/AlGaAs coupled quantum well structure. The spin texture is
observed around the exciton rings. The observed phenomena include: a ring of linear polarization, a vortex of linear polarization with polarization
perpendicular to the radial direction, an anisotropy in the exciton flux, a skew of the exciton fluxes in orthogonal circular polarizations and a
corresponding four-leaf pattern of circular polarization, a periodic spin
texture, and extended exciton coherence in the region of the polarization vortex. The data indicate a transport regime where the spin polarization is locked to the direction of particle propagation and scattering is suppressed.
\end{abstract}

\maketitle

Spin ordering forms the basis for a number of fundamental effects in
physics. Macroscopic spin ordering with neighboring spins parallel or
antiparallel to each other is responsible for ferromagnetism and
antiferromagnetism. More complex spin structures, also known as spin
textures, determine the physical properties of exotic states of matter and are currently under intense investigation. Those spin structures include
skyrmion spin textures in quantum Hall ferromagnets \cite{Girvin00} and
graphene \cite{Fertig07}, ferromagnetic domains and spin vortices in atom
Bose-Einstein condensates \cite{Sadler06}, skyrmion lattices in chiral
magnets \cite{Muhlbauer09,Yu10}, and half-vortices in He3 \cite{Salomaa87} and polariton condensates \cite{Lagoudakis09}. Spin textures are also
characteristic of topological insulators \cite{Qi10}.

\begin{figure}[tbp]
\includegraphics[width=8.5cm]{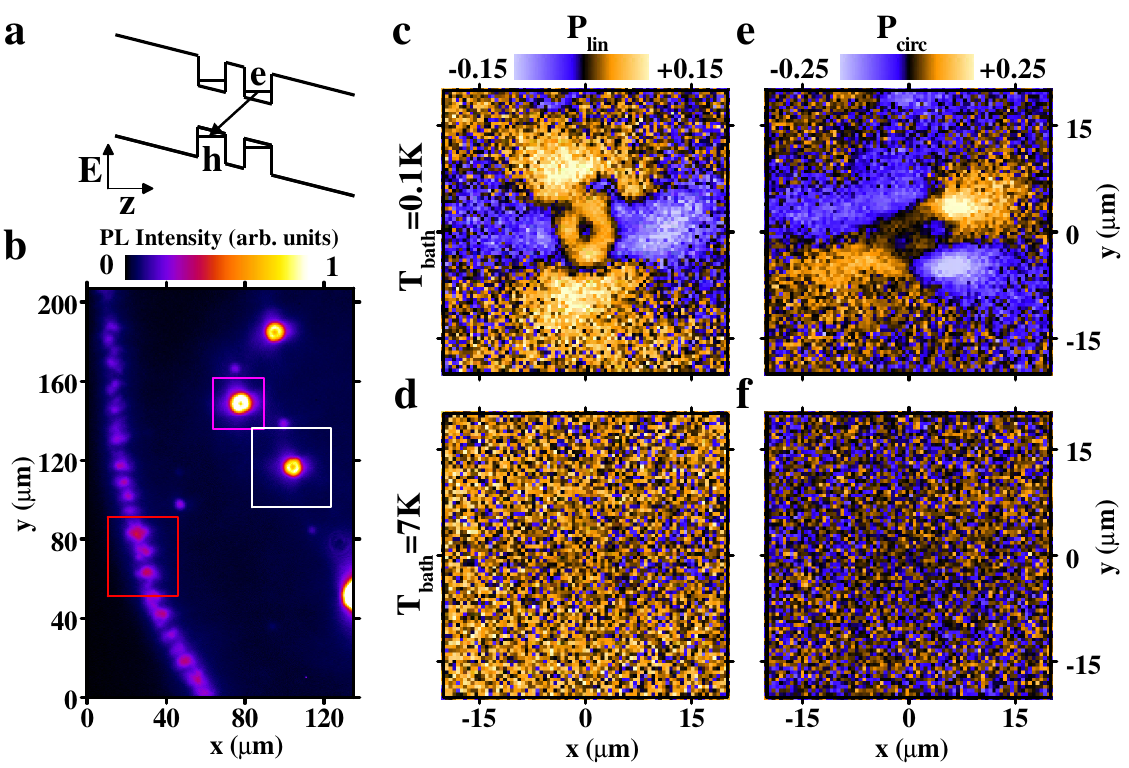}
\caption{Spin texture around LBS. (a) Energy-band diagram of the
CQW; e, electron; h, hole. (b) A segment of the emission pattern of indirect excitons. The white (red) box indicates the LBS (external ring) region presented in Fig. 1-3 (4). (c, d) Texture of linear polarization of the emission of indirect excitons $P_{lin} = (I_x - I_y) / (I_x + I_y)$ around LBS at (c) $T_{bath} = 0.1$ K and (d) $T_{bath} = 7$ K. (e, f) Texture of circular polarization of the emission of indirect excitons $P_{\protect\sigma} = (I_{\protect\sigma^+} - I_{\protect\sigma^-}) / (I_{\protect\sigma^+} + I_{\protect\sigma^-})$ around LBS at (e) $T_{bath} = 0.1$ K and (f) $T_{bath} = 7$ K.}
\end{figure}

Here, we report on the observation of a spin texture in a cold exciton gas. The spin texture appears when the exciton gas is cooled below a few Kelvin. The cold exciton gas is realized in a GaAs/AlGaAs coupled quantum well structure (CQW), Fig. 1a. Indirect excitons in CQW are characterized by properties, which are essential for the observation of spin textures: (i) Long lifetimes of indirect excitons allow them to cool to low temperatures within about 0.1 K of the lattice temperature, which can be lowered to well below 1 K in a dilution refrigerator. This allows the realization of a cold exciton gas with temperature well below the temperature of quantum degeneracy, which is in the range of a few Kelvin for typical exciton densities $\sim 10^{10}$ cm$^{-2}$ for the studied GaAs/AlGaAs CQW \cite{Butov01}. (ii) Long lifetimes of indirect excitons allow them to travel over large distances before recombination \cite{Hagn95,Larionov00,Butov02,Gartner06,Ivanov06}. (iii) The exciton spin relaxation induced by the electron-hole exchange interaction is strongly suppressed for indirect excitons with small electron-hole overlap and, as a result, spin relaxation times of indirect excitons can be orders of magnitude longer than those of regular excitons \cite{Vinattieri94,Larionov08,Leonard09,Kowalik-Seidl10}.

The spin texture is observed around the rings in the exciton emission
pattern. Figure 1b shows a segment of the exciton pattern with a section of the external ring and smaller localized bright spot (LBS) rings. Such
exciton rings were observed a few years ago \cite{Butov02} and studied in
\cite{Butov04,Rapaport04,Chen05,Haque06,Yang10}. Both the external rings and LBS rings form exciton sources; excitons are generated inside the rings at the ring shaped interface between electron-rich and hole-rich regions. The former is created by current through the structure (specifically, by the current filament at the LBS center in the case of the LBS ring), while the latter is created by photoexcitation \cite{Butov04,Rapaport04,Chen05,Haque06,Yang10}. The indirect excitons cool down to the lattice temperature within a few microns from the source \cite{Hammack09} so, in the case of indirect excitons, the rings serve as sources of cold excitons.

\begin{figure}[tbp]
\includegraphics[width=7.5cm]{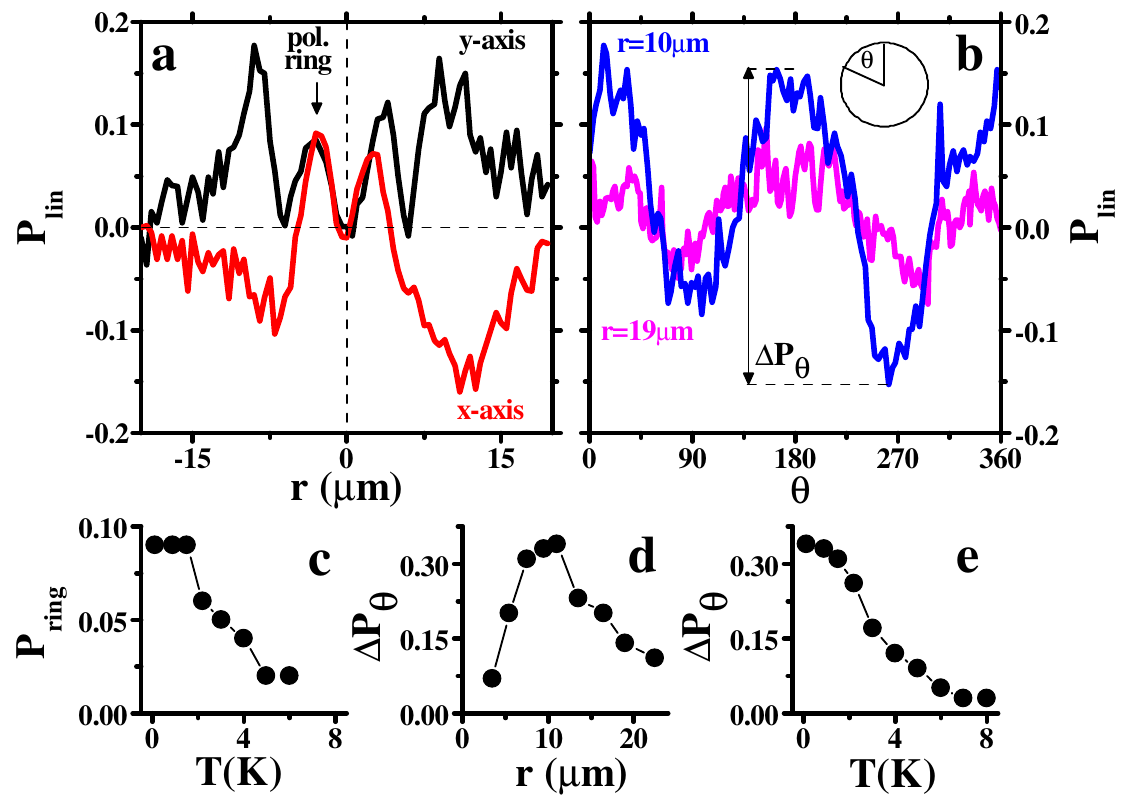}
\caption{Texture of linear polarization around LBS. (a) The radial variation of $P_{lin}$ along $x$ (red) and $y$ (black) at $T_{bath} = 0.1$ K. (b) The azimuthal variation of $P_{lin}$ at the distance from the LBS center $r = 10 \protect\mu$m (blue) and $r = 19 \protect\mu$m (magenta) at $T_{bath} = 0.1$ K. (c) The maximum $P_{lin}$ in the $x$-polarization ring vs. temperature. (d) The amplitude of azimuthal variation of $P_{lin}$, $\Delta P_{\protect\theta}$, [see (b)] at $T_{bath} = 0.1$ K vs. $r$. (e) $\Delta P_{\protect\theta}$ at $r = 10 \protect\mu$m vs. temperature.}
\end{figure}

In our experiments, the photoexcitation is $>400$ meV above the energy of
indirect excitons and the $10\mu$m-wide excitation spot is $>80\mu$m away from both the LBS and external ring. This eliminates the influence of the
laser polarization on the spin pattern. In contrast to the earlier
experiments \cite{Leonard09} where excitons were generated in the excitation spot, in the present experiments, the excitons are generated in the exciton rings. Such generation does not suffer from the laser-induced heating and results in the creation of cold indirect excitons \cite{Butov04}. The achieved low temperatures are essential for the observation of exciton spin textures as detailed below. Experiments are performed in an optical dilution refrigerator. The spin texture is revealed in the polarization texture of the emitted light and is measured by polarization-resolved imaging. We also measured exciton coherence by shift-interferometry. The interfering emission images produced by arm 1 and 2 of the Mach-Zehnder (MZ) interferometer are laterally shifted relative to each other to measure the interference between the emission of excitons, which are spatially separated by $\delta_{s}$. This method allows direct measurement of the exciton coherence length $\xi$ in the QW plane. $\xi$ is evaluated here as $\delta_{s}$ at which the interference visibility drops by $e$ times. Details on the experimental setup and the CQW structure are given in the supplementary materials.

\begin{figure}[tbp]
\includegraphics[width=7.5cm]{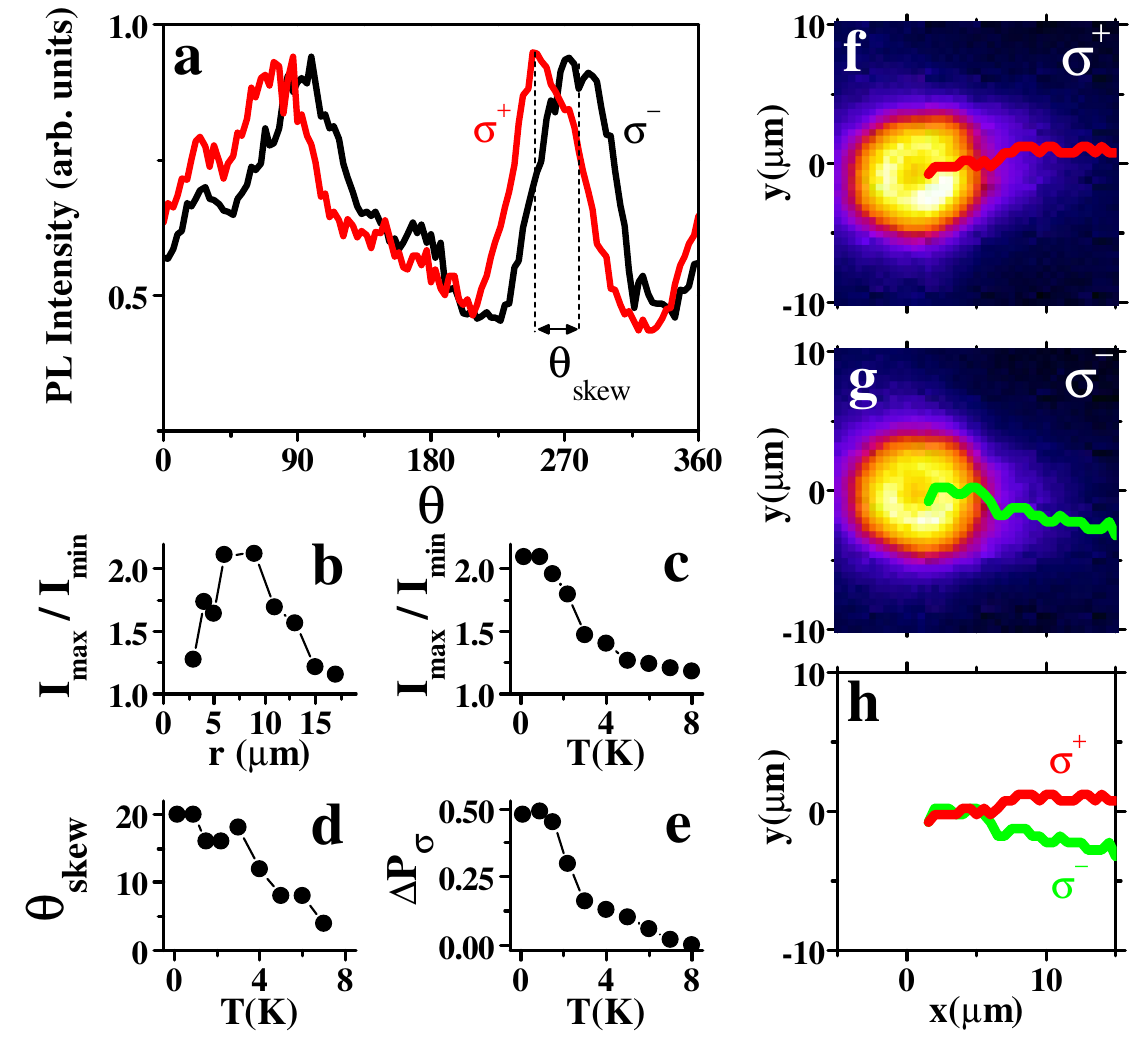}
\caption{Texture of circular polarization around LBS. (a) Azimuthal variation of the emission intensity of indirect excitons at $r = 8 \protect\mu$m in $\protect\sigma^+$ (red) and $\protect\sigma^-$ (black) polarizations. (b) The ratio of maximum to minimum in the azimuthal variation of the total emission intensity of indirect excitons $I_{max} / I_{min}$ at $T_{bath} = 0.1 $ K vs. $r$. (c) $I_{max}/I_{min}$ at $r = 8 \protect\mu$m vs. temperature. (d) A skew of the emission peaks in $\protect\sigma^+$ and $\protect\sigma^-$ polarizations $\protect\theta_{skew}$ around $r = 8 \protect\mu$m [see (a)] vs. temperature. (e) The amplitude of variation of $P_{\protect\sigma}$ in the LBS area shown in Fig. 1e,f vs. temperature. (f-h) A trace of $\protect\sigma^+$ (red) and $\protect\sigma^-$ (green) emission peak around $\protect\theta = 270^{\circ}$ [see (a)]; An emission image in (f) $\protect\sigma^+$ and (g) $\protect\sigma^-$ polarization is also shown.}
\end{figure}

The observed phenomena are qualitatively similar for both exciton sources -- the external ring and LBS ring, as is detailed below. An LBS ring is close to a model radially symmetric source of excitons; therefore, we first discuss the features of the spin texture around the LBS. All LBS rings in the emission pattern show similar spin textures. The observed phenomena are listed below.

1. A ring of linear polarization is observed around the LBS center (Fig. 1c, 2a). Measurements with a rotating polarizer confirm polarization along the $x$-direction. The ring radius, where the linear polarization is maximum, is about $3 \mu$m. The ring of linear polarization vanishes with increasing temperature (Fig. 1d, 2c).

2. A vortex of linear polarization is observed around the LBS center (Fig. 1c, 2a,b). Figures 1c, 2a,b show that $y$-polarization is observed along $x$-direction and $x$-polarization is observed along $y$-direction. Measurements with a rotating polarizer confirm that the polarization is perpendicular to the radial direction for all azimuthal angles $\theta$ for the polarization vortex. The polarization vortex is observed in the range of distances from the LBS center $3 \lesssim r \lesssim 20 \mu$m (Fig. 1c, 2a,d). The ring of linear polarization and polarization vortex overlap at $r = 3 - 5 \mu$m forming a more complex spin texture. The polarization vortex vanishes with increasing temperature (Fig. 1d, 2e).

3. The flux of excitons from the LBS origin is anisotropic (Fig. 3a): The
emission intensity is enhanced along the $x$-axis, the polarization
direction in the polarization ring. The exciton flux anisotropy is observed in the range of distances from the LBS center $3 \lesssim r \lesssim 15 \mu$m (Fig. 3b) and vanishes with increasing temperature (Fig. 3c) so the exciton flux from the LBS origin is isotropic at high temperatures.

4. A skew of the exciton fluxes in orthogonal circular polarizations (Fig. 3a, f-h) and a corresponding four-leaf pattern of circular polarization (Fig. 1e) is observed around the LBS. The skew of the exciton flux is observed in the range of distances from the LBS center $5 \lesssim r \lesssim 15 \mu$m (Fig. 3f-h). The skew of the exciton flux and the corresponding four-leaf pattern of circular polarization vanish with increasing temperature (Fig. 1f, 3d,e).

\begin{figure}[tbp]
\centering
\includegraphics[width=8cm]{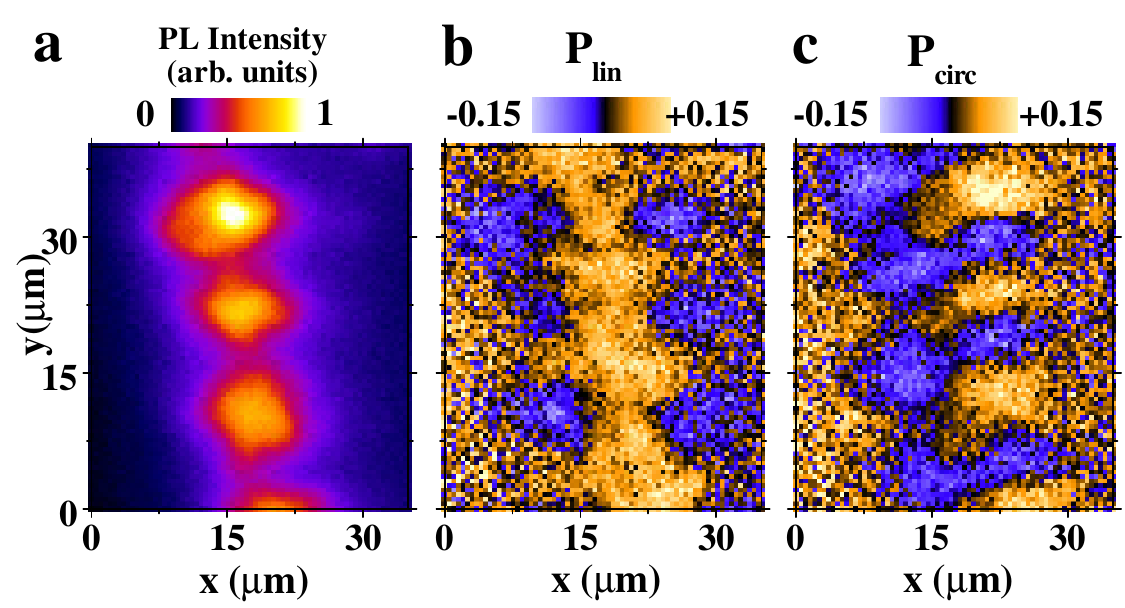}
\caption{Spin texture around external ring. (a) Emission of
indirect excitons in the external ring region. (b, c) Pattern of linear (b) and circular (c) polarization around the external ring. $T_{bath} = 0.1$ K.}
\end{figure}

5. Similar polarization textures are observed in the external ring region
(Fig. 4). At low temperatures, the macroscopically ordered exciton state
(MOES) characterized by a spatially ordered array of higher-density beads
\cite{Butov02} and a large exciton coherence length \cite{Yang06} forms in the external ring. The polarization texture in the external ring region
appears as the superposition of the polarization textures produced by the
MOES beads with each being similar to the texture produced by an LBS
(compare Fig. 4b with Fig. 1c and Fig. 4c with Fig. 1e). A periodic array of beads in the MOES (Fig. 4a) creates a periodic polarization texture (Fig. 4b,c). Both the position of the external ring and the wavelength of the exciton density wave are controlled by the laser excitation indicating that the exciton density modulation in the MOES is not governed by disorder in the sample. The presence of spin texture in the external ring region confirms that the spin texture does not arise due to a local defect structure.

6. An extended exciton coherence with a large coherence length $\xi$ is observed in the region of the polarization vortex (Fig. 5a). In contrast, $\xi$ is short in the region of the polarization ring, which is close to the hot LBS center. With reducing temperature, the exciton spin textures emerge in concert with coherence (Fig. 2e and 5b).

\begin{figure}[tbp]
\centering
\includegraphics[width=7.5cm]{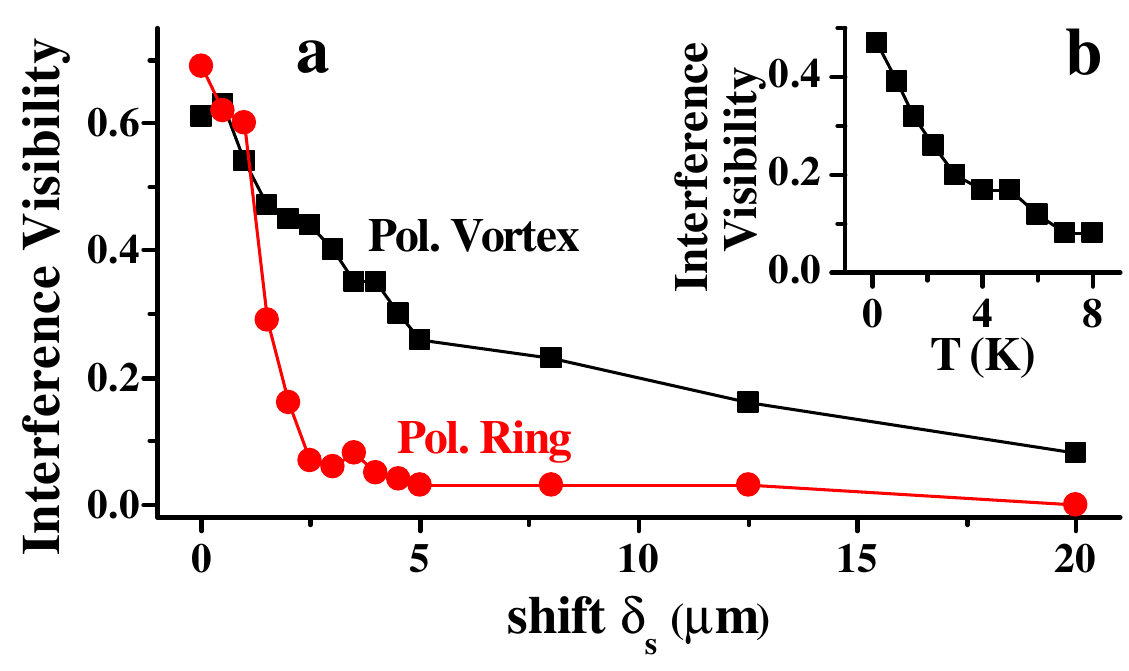}
\caption{Exciton coherence measured by shift-interferometry. The interfering emission images produced by arm 1 and 2 of the MZ interferometer are shifted relative to each other by $\protect\delta$ along $x$ direction to measure the interference between the emission of excitons, which are laterally separated by $\protect\delta_s = \protect\delta / M$, where $M=25$ is the image magnification. (a) Interference visibility $V$ vs. shift $\protect\delta_s$ for the polarization vortex (squares), $10 \protect\mu$m left of LBS center, and the polarization ring (circles), $2 \protect\mu$m left of LBS center (the LBS is in magenta box in Fig. 1b). $T_{bath} = 0.1$ K. (b) $V$ in the polarization vortex vs. temperature for $\protect\delta_s = 2\protect\mu$m.}
\end{figure}

Below, we briefly discuss the observed phenomena and consider a model which can lead to the appearance of an exciton spin texture. This model is based on ballistic exciton transport and precession of spins of electrons and holes. The heavy hole excitons in GaAs QWs may have four spin projections on the structure axis: $-2,-1,+1,+2$. The spin states $-1$ and $+1$ are optically active and contribute to left- and right-circularly polarized emission, while the states $-2$ and $+2$ are so called dark excitons which are optically inactive \cite{Maialle93,Silva97,Wu08}. The splitting of linearly polarized exciton states originates from in-plane anisotropy induced by the crystallographic axis orientation and strain. The precession of spins of electrons and holes moving in the QW plane originates from this splitting and spin-orbit interaction which is described by a Dresselhaus Hamiltonian $H_{e}=\beta_{e}\left( k_{x}^{e}\sigma_{x}-k_{y}^{e}\sigma_{y}\right)$ for electrons and $H_{h}=\beta _{h}\left( k_{x}^{h}\sigma_{x}+k_{y}^{h}\sigma_{y}\right)$ for holes \cite{Luo10}, where $\mathbf{k}_{e,h}$ are electron and hole wave-vectors, $\beta _{e,h}$ are constants, and $\sigma _{x,y}$ are Pauli matrices.

The appearance of an exciton spin texture within this model is outlined below. Details are presented in the supplementary materials. {\it A ring of linear polarization}: The splitting between the linearly polarized exciton states leads to linear polarization of the exciton gas at thermal equilibrium. Heating of the exciton gas at the origin reduces the polarization degree in the center and, as a result, leads to the appearance of a ring of linear polarization. {\it A vortex of linear polarization}: Spin-orbit coupling provides two channels for conversion of bright to dark excitons and vice versa with the phase gained by excitons during such conversion dependent on their propagation direction. This results in populating the exciton states whose polarization is normal to the wave vector and, in turn, the appearance of a vortex of linear polarization. {\it A four-leaf pattern of circular polarization}: The beats between orthogonal linearly polarized exciton states lead to the appearance of a four-leaf pattern of circular polarization in a qualitative similarity to the optical spin Hall effect \cite{Leyder07}. The appearance of a vortex of linear polarization and four-leaf pattern of circular polarization within the model rely on ballistic propagation of excitons with coherent spin precession.

The data are discussed below. {\it A ring of linear polarization}: The current filament forms a heating source at the LBS center while no such local heating source is present in the external ring \cite{Butov04}, consistent with the ring of linear polarization in the LBS area and its absence in the external ring area (Fig. 1c and 4b).

{\it A vortex of linear polarization}: The appearance of this spin texture indicates a transport regime where the spin polarization is locked to the direction of particle propagation, scattering is suppressed, and the mean free path is dramatically enhanced. Indeed, the vortex of linear polarization indicates that the exciton polarization is perpendicular to the direction of exciton propagation from the origin. The observation of the polarization vortex up to $\sim 20 \mu$m away from the origin (Fig. 2d) indicates that the exciton propagation direction is maintained over this distance. Such propagation requires suppression of exciton scattering on disorder. In comparison, for a classical exciton transport even with a very high diffusion coefficient $D=100$ cm$^{2}$/s, the exciton mean free path is only $\sim 1 \mu$m so maintaining polarization normal to the direction to the origin over $\sim 20 \mu$m is unlikely due to multiple collisions.

{\it Extended coherence}: The suppression of exciton scattering in the region of the polarization vortex is evidenced by extended exciton coherence with a large coherence length $\xi \sim 8\mu$m (Fig. 5a). In comparison, for a classical exciton gas, $\xi_{cl}$ is close to the thermal de Broglie wavelength, which scales $\propto T^{-1/2}$ and is about $0.5 \mu$m at 0.1 K. Large $\xi \gg \xi_{cl}$ in the region of the polarization vortex indicates a coherent exciton state with a much narrower than classical exciton distribution in momentum space, characteristic of a condensate.

We thank Misha Fogler, Jorge Hirsch, Leonid Levitov,
Yuriy Rubo, Lu Sham, Ben Simons, Masha Vladimirova, and Congjun Wu for discussions. This work was supported by the DOE Office of Basic Energy Sciences under award DE-FG02-07ER46449. The
coherence measurements and development of spectroscopy in the dilution
refrigerator were also supported by ARO under award W911NF-08-1-0341 and NSF under award 0907349. T.O. was supported by the Ministry of Education and the Grant Agency of the Czech Republic through the grants MSM0021620834 and P204/10/P326.


\begin{thebibliography}{99}
\bibitem{Girvin00}
S.M. Girvin, Physics Today 39 (June 2000).

\bibitem{Fertig07}
H.A. Fertig, L. Brey, Eur. Phys. J. \textbf{148}, 143 (2007).

\bibitem{Sadler06}
L.E. Sadler, J.M. Higbie, S.R. Leslie, M. Vengalattore, D.M. Stamper-Kurn, Nature \textbf{443}, 312 (2006).

\bibitem{Muhlbauer09}
S.M\"{u}hlbauer, B. Binz, F. Jonietz, C. Pfleiderer, A. Rosch, A. Neubauer, R. Georgii, P. B\"{o}ni, Science \textbf{323}, 915 (2009).

\bibitem{Yu10}
X.Z. Yu, Y. Onose, N. Kanazawa, J.H. Park, J.H. Han, Y. Matsui, N. Nagaosa, Y. Tokura, Nature \textbf{465}, 901 (2010).

\bibitem{Salomaa87}
M.M. Salomaa, G.E. Volovik, Rev. Mod. Phys. \textbf{59}, 533 (1987).

\bibitem{Lagoudakis09}
K.G. Lagoudakis, T. Ostatnick\'{y}, A.V. Kavokin, Y.G. Rubo, R. Andr\'{e}, B. Deveaud-Pl\'{e}dran, Science \textbf{326}, 974 (2009).

\bibitem{Qi10}
Xiao-Liang Qi, Shou-Cheng Zhang, Physics Today 33 (Jan 2010).

\bibitem{Butov01}
L.V. Butov, A.L. Ivanov, A. Imamoglu, P.B. Littlewood, A.A. Shashkin, V.T. Dolgopolov, K.L. Campman, A.C. Gossard, Phys. Rev. Lett. \textbf{86}, 5608 (2001).

\bibitem{Hagn95}
M. Hagn, A. Zrenner, G. B{\"o}hm, G. Weimann, Appl. Phys. Lett. \textbf{67}, 232 (1995).

\bibitem{Larionov00}
A.V. Larionov, V.B. Timofeev, J. Hvam, K. Soerensen, JETP \textbf{90}, 1093 (2000).

\bibitem{Butov02}
L.V. Butov, A.C. Gossard, D.S. Chemla, Nature \textbf{418}, 751 (2002).

\bibitem{Gartner06}
A. Gartner, A.W. Holleithner, J.P. Kotthaus, D. Schul, Appl. Phys. Lett. \textbf{89}, 052108 (2006).

\bibitem{Ivanov06}
A.L. Ivanov, L.E. Smallwood, A.T. Hammack, S. Yang, L.V. Butov, A.C. Gossard, Europhys. Lett. \textbf{73}, 920 (2006).

\bibitem{Vinattieri94}
A. Vinattieri, J. Shah, T.C. Damen, D.S. Kim, L.N. Pfeiffer, M.Z. Maialle, L.J. Sham, Phys. Rev. B \textbf{50}, 10868 (1994).

\bibitem{Larionov08}
A.V. Larionov, L.E. Golub, Phys. Rev. B \textbf{78}, 033302 (2008).

\bibitem{Leonard09}
J.R. Leonard, Y.Y. Kuznetsova, Sen Yang, L.V. Butov, T. Ostatnick\'{y}, A. Kavokin, A.C. Gossard, Nano Lett. \textbf{9}, 4204 (2009).

\bibitem{Kowalik-Seidl10}
K. Kowalik-Seidl, X.P. V{\"o}gele, B.N. Rimpfl, S. Manus, J.P. Kotthaus, D. Schuh, W. Wegscheider, A.W. Holleitner, Appl. Phys. Lett. \textbf{97}, 011104 (2010).

\bibitem{Butov04}
L.V. Butov, L.S. Levitov, A.V. Mintsev, B.D. Simons, A.C. Gossard, D.S. Chemla, Phys. Rev. Lett. \textbf{92}, 117404 (2004).

\bibitem{Rapaport04}
R. Rapaport, G. Chen, D. Snoke, S.H. Simon, L. Pfeiffer, K. West, Y. Liu, S. Denev, Phys. Rev. Lett. \textbf{92}, 117405 (2004).

\bibitem{Chen05}
G. Chen, R. Rapaport, S.H. Simon, L. Pfeiffer, K. West, Phys. Rev. B \textbf{71}, 041301 (2005).

\bibitem{Haque06}
M. Haque, Phys. Rev. E \textbf{73}, 066207 (2006).

\bibitem{Yang10}
Sen Yang, L.V. Butov, L.S. Levitov, B.D. Simons, A.C. Gossard, Phys. Rev. B \textbf{81}, 115320 (2010).

\bibitem{Hammack09}
A.T. Hammack, L.V. Butov, J. Wilkes, L. Mouchliadis, E.A. Muljarov, A.L. Ivanov, A.C. Gossard, Phys. Rev. B \textbf{80}, 155331 (2009).

\bibitem{Yang06}
Sen Yang, A.T. Hammack, M.M. Fogler, L.V. Butov, A.C. Gossard, Phys. Rev. Lett. \textbf{97}, 187402 (2006).

\bibitem{Maialle93}
M.Z. Maialle, E. A. de Andrada e Silva, L.J. Sham, Phys. Rev. B \textbf{47}, 15776 (1993).

\bibitem{Silva97}
E.A. de Andrada e Silva, G.C. La Rocca, Phys. Rev. B \textbf{56}, 9259 (1997).

\bibitem{Wu08}
Congjun Wu, Ian Mondragon-Shem, arXiv:0809.3532v3

\bibitem{Luo10}
Jun-Wei Luo, A.N. Chantis, M. van Schilfgaarde, G. Bester, A. Zunger, Phys. Rev. Lett. \textbf{104}, 066405 (2010).

\bibitem{Leyder07}
C. Leyder, M. Romanelli, J.Ph. Karr, E. Giacobino, T.C.H. Liew, M.M. Glazov, A.V. Kavokin, G. Malpuech, A. Bramati, Nature Physics
\textbf{3}, 628 (2007).

\end{thebibliography}
\end{document}

\centerline{\bf SUPPLEMENTARY MATERIALS} \vskip5mm

\textbf{Model of spin texture around LBS}

The model considers excitons around an LBS ring, which is close to a
radially symmetric source of excitons, for details on LBS rings see
[12,19,23]. An LBS ring has a hot core region [19], which is not described
by the model. The initial exciton state considered by this model is a ring
around the LBS center where the exciton temperature reaches the limit below
which excitons propagate ballistically and keep coherence
within the limits of the observed spin texture. The model assumes a thermal
distribution of excitons on this ring and their coherent (ballistic) radial
propagation away from the center with velocity $v=\hbar k_{ex}/m$, where $%
\mathbf{k}_{ex}=\mathbf{k}_{e}+\mathbf{k}_{h}$ is the exciton wave vector
and $m$ is the exciton effective mass in the plane of the structure, $%
\mathbf{k}_{e}=\frac{m_{e}}{m}\mathbf{k}_{ex},$ $\mathbf{k}_{h}=\frac{m_{h}}{%
m}\mathbf{k}_{ex}$, $m_{e}$ and $m_{h}$ are electron and hole in-plane
effective masses, respectively. In the basis of four exciton states with
spins $\{{+}1,{-}1,{+}2,{-}2\}$, the coherent spin dynamics in the system is
governed by a model matrix Hamiltonian: 
\begin{equation}
H_{exc}=\left( 
\begin{array}{cccc}
E_{bright} & {-}\delta _{b} & k_{e}\beta _{e}e^{-i\varphi } & k_{h}\beta
_{h}e^{-i\varphi } \\ 
{-}\delta _{b} & E_{bright} & k_{h}\beta _{h}e^{i\varphi } & k_{e}\beta
_{e}e^{i\varphi } \\ 
k_{e}\beta _{e}e^{i\varphi } & k_{h}\beta _{h}e^{-i\varphi } & E_{dark} & {-}%
\delta _{d} \\ 
k_{h}\beta _{h}e^{i\varphi } & k_{e}\beta _{e}e^{-i\varphi } & {-}\delta _{d}
& E_{dark}%
\end{array}%
\right) ,
\end{equation}%
where $E_{bright}$ \ and $E_{dark}$ are energies of bright and dark excitons
in an ideal isotropic QW, $\delta _{b}$ and $\delta _{d}$ describe the
effect of in-plane anisotropy induced by the crystallographic axis
orientation and strain and resulting in the splitting of exciton states
linearly polarized along $[110]$ and $[0-10]$ axes of symmetry. The angle $%
\varphi $ is defined relative to the $x\equiv \lbrack 110]$ axis. This
Hamiltonian has been constructed from the spin--orbit Hamiltonians for
electrons and holes [29], including an exciton spin anisotropy: 
\begin{equation}
H_{exc}=H_{anis}+\mathbf{1}_{e}\otimes H_{h}+H_{e}\otimes \mathbf{1}_{h}\,,
\end{equation}%
where the operators $\mathbf{1}_{e,h}$ are 2$\times $2 identity matrices, $%
H_{e}=\beta _{e}\left( k_{x}^{e}\sigma _{x}-k_{y}^{e}\sigma _{y}\right) $
and $H_{h}=\beta _{h}\left( k_{x}^{h}\sigma _{x}+k_{y}^{h}\sigma _{y}\right)
,$

\begin{equation}
H_{anis}=\left( 
\begin{array}{cccc}
E_{bright} & {-}\delta _{b} & 0 & 0 \\ 
{-}\delta _{b} & E_{bright} & 0 & 0 \\ 
0 & 0 & E_{dark} & {-}\delta _{d} \\ 
0 & 0 & {-}\delta _{d} & E_{dark}%
\end{array}%
\right) .
\end{equation}%
The evolution of the density matrix is then described in terms of the
quantum Liouville equation: 
\begin{equation}
i\hbar \,d\rho /dt=\big[H_{exc},\rho \big]
\end{equation}
The initial condition in the simulations is the thermal density matrix on
the low--temperature ring around the LBS core: 
\begin{equation}
\rho (R=R_{0})=\left( 
\begin{array}{cccc}
(n_{1}+n_{2})/2 & (n_{1}-n_{2})/2 & 0 & 0 \\ 
(n_{1}-n_{2})/2 & (n_{1}+n_{2})/2 & 0 & 0 \\ 
0 & 0 & (n_{3}+n_{4})/2 & (n_{3}-n_{4})/2 \\ 
0 & 0 & (n_{3}-n_{4})/2 & (n_{3}+n_{4})/2%
\end{array}%
\right) \,,
\end{equation}%
where the populations of the linearly polarized bright states are $%
n_{1,2}=\exp \big[-\big(E_{bright}\mp \delta _{b}\big)/k_{B}T\big]$ and
populations of the dark states are $n_{3,4}=\exp \big[-\big(E_{dark}\mp
\delta _{d}\big)/k_{B}T\big]$. Symbol $k_{B}$ stands for the Boltzmann
constant. The patterns of circular and linear polarizations are extracted
from the density matrix using the formulas $P_{\sigma }(R)=\big[\rho
_{11}(R)-\rho _{22}(R)\big]/\big[\rho _{11}(R)+\rho _{22}(R)\big]$ and $%
P_{lin}(R)=2\mathrm{Re}\big[\rho _{12}(R)\big]/\big[\rho _{11}(R)+\rho
_{22}(R)\big]$, respectively.

Figures S1a,b show the results of our calculations, considering the
following parameters. The energies of exciton states were defined by the
parameters $E_{dark}=0\ \mu $eV, $E_{bright}=5\ \mu $eV, $\delta _{b}={+}1\
\mu $eV, $\delta _{d}={-}8\ \mu $eV. The
Dresselhaus coefficients were taken to be $\beta _{e}=25\mathrm{\ meV%
\mathring{A}}$ and $\beta _{h}=23.5\mathrm{\ meV\mathring{A}}$, the first
value has been experimentally measured in our earlier work [17], the value
for $\beta _{h}$ is in qualitative agreement with the calculations in [29]. The
exciton temperature is $T=0.2$~K, exciton propagation velocity $v=3.4\cdot
10^{5}$~cm/s, effective masses $m=0.23\ m_{0}$, $m_{e}=0.07\ m_{0}$, $%
m_{h}=0.16\ m_{0}$ with $m_{0}$ being free electron mass, and $R_{0}=4.8\ \mu 
$m.

For a detailed fit of the experiment the spin dependent exciton-exciton
interactions may need to be taken into account. Nevertheless, the present
(linear) theory reproduces all main features of the spin texture observed in
the experiment, namely, a ring of linear polarization (Fig. S1a), a vortex
of linear polarization with the polarization perpendicular to the radial
direction (Fig. S1a), and a four-leaf pattern of circular polarization (Fig.
S1b).

The model is based on the assumption of ballistic radial transport of
excitons. In this regime, the main mechanism of hole spin relaxation
(Elliott-Yafet mechanism linked with the scattering of holes by impurities)
is suppressed. The spin dynamics of electrons and holes is governed by the
spin-orbit interaction described by the Dresselhaus Hamiltonian, which we
use in the model. The splitting of linearly polarized states of bright and
dark excitons is an essential element of the model, which allows to describe
the central linear polarization ring and the build-up of circular
polarization. Such splittings typically appear in MBE grown samples due to
monolayer fluctuations of the heteroboundaries. The values of these two
splittings in our sample are fitting parameters of the model. A good qualitative agreement
between the model and experiment indicates that the observed spin texture is a
characteristic feature of a new exciton transport regime where coherence is
kept over large distances.

\begin{figure}[tbp]
\includegraphics[width=0.4\textwidth]{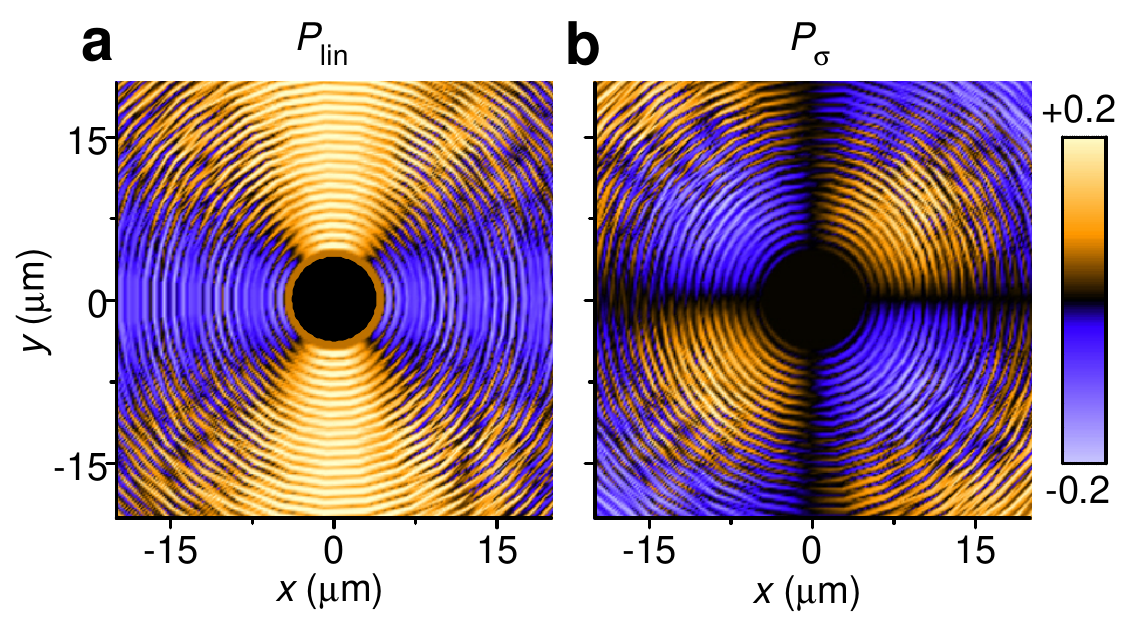}
\caption{(a,b) Simulated (a) pattern of linear polarization, and (b) pattern of circular
polarization around LBS. The LBS center is in the origin. The black region
in the center refers to a hot exciton gas, which is not considered in the
simulations.}
\end{figure}

\vskip5mm \textbf{CQW structure and experimental setup}

Experiments were performed on a $n^+ - i - n^+$ GaAs/AlGaAs CQW structure
grown by molecular-beam epitaxy. The $i$ region consists of a single pair of
8 nm GaAs QWs separated by a 4 nm Al$_{0.33}$Ga$_{0.67}$As barrier and
surrounded by 200 nm Al$_{0.33}$Ga$_{0.67}$As barrier layers. The $n^+$
layers are Si-doped GaAs with $N_{Si} = 5 \times 10^{17} cm^{-3}$. The
electric field in the $z$ direction is controlled by the external gate
voltage applied between $n^+$ layers. The small in-plane disorder in the CQW
is indicated by the photoluminescence linewidth $\approx 1$ meV. The laser
excitation is performed by a HeNe laser at $\lambda = 633$ nm. The x-polarization is along the sample cleavage direction within the experimental accuracy.

\begin{figure*}[h]
\centering
\includegraphics[width=0.75\textwidth]{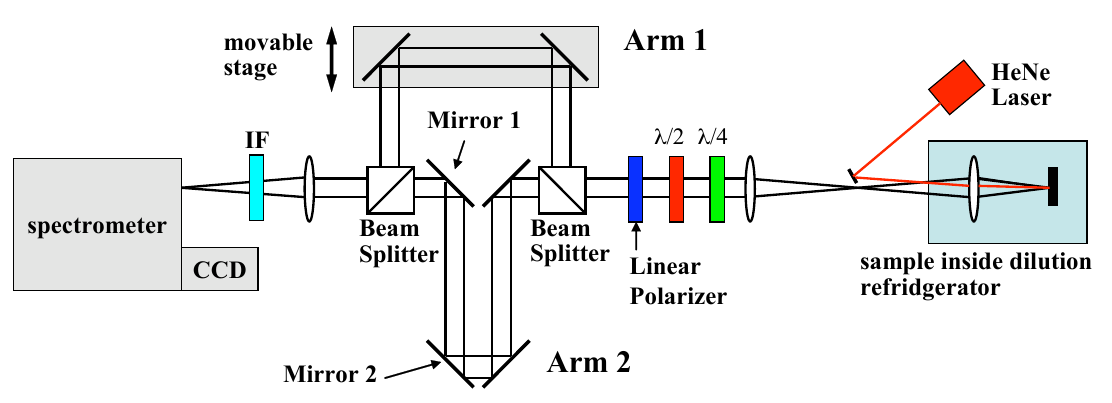}
\caption{Schematic of interferometric setup.}
\end{figure*}

We use a Mach-Zehnder (MZ) interferometer to probe the coherence of the
exciton system (Fig. S2). The emission beam is made parallel by an objective
inside the optical dilution refrigerator and lenses. A combination of a
quarter-wave plate ($\lambda /4$) and a half-wave plate ($\lambda /2$)
converts the measured polarization of the emission to $y-$polarization,
which is then selected by a linear polarizer. This ensures only $y-$%
polarized light enters the MZ interferometer thus eliminating
polarization-dependent effects, which otherwise can be caused by the
interferometer and spectrometer. The emission is then split between arm 1
and arm 2 of the MZ interferometer. The path lengths of arm 1 and arm 2 are
set equal using the movable stage. The interfering emission images produced
by arm 1 and 2 of the MZ interferometer are shifted relative to each other
by $\delta$ along $x$ direction to measure the interference between the
emission of excitons, which are laterally separated by $\delta_s = \delta /
M $, where $M=25$ is the image magnification. Mirror 1 and Mirror 2 are used
to control the fringe period, and Mirror 2 is used to control $\delta_s$.
After the interferometer, the emission is filtered by an $800 \pm 5$nm
interference filter (IF), which is adjusted to the emission wavelength of
indirect excitons, and focused to produce an emission image of indirect
excitons. The latter is recorded by a liquid-nitrogen cooled CCD. We collect
emission for arm 1 open, arm 2 open, and both arms open and obtain $%
I_{interf} = (I_{12} - I_1 - I_2)/(2 \sqrt{I_1 I_2})$, where $I_1, I_2,$ and 
$I_{12}$ is the emission intensity of indirect excitons for arm 1, arm 2,
and both arms of the MZ interferometer, respectively. Here we use the
general equation for interference between two partially coherent sources at $%
r_1$ and $r_2$: $I_{12}= I_1 + I_2 + 2 \sqrt{I_1I_2}\cos[\delta \phi(r_1,r_2)%
]\zeta(r_1,r_2)$, where $\delta \phi(r_1,r_2)$ is the phase difference
between $r_1$ and $r_2$ and $\zeta(r_1,r_2)$ is coherence between $r_1$ and $%
r_2$ [see e.g. P.W. Milonni, J.H. Eberly, Lasers (Wiley, New York, 1988),
chap. 15]. Then, $I_{interf}=\cos[\delta \phi(r_1,r_2)] \zeta(r_1,r_2)$,
which contains the relevant quantities of phase difference and coherence.
For our experimental geometry, $\cos[\delta \phi(r_1,r_2)]$ oscillates along
the $y$-axis, and coherence $\zeta(r_1,r_2)$ for $r_1 - r_2 = \delta_s$ is
given by the amplitude of these oscillations.